\documentclass[twocolumn,prl,aps,10pt,footinbib,noshowpacs]{revtex4-1}
\usepackage{graphicx}
\usepackage[utf8x]{inputenc}
\usepackage{amsmath}
\usepackage{lineno}
\usepackage{color}

\begin{document}

\title{Fermi blockade of the electron-phonon interaction:
why strong coupling effects may not be seen in optimally doped high temperature superconductors.}

\author{Andrey S. Mishchenko$^{1,2}$, Naoto Nagaosa$^{1,3}$,  and Nikolay Prokof'ev$^{4,2}$
}

\affiliation{
$^1$RIKEN Center for Emergent Matter Science (CEMS), Wako, Saitama 351-0198, Japan\\
$^2$NRC ``{\it Kurchatov Institute}'', 123182 Moscow, Russia\\
$^3$Department of Applied Physics, The University of Tokyo 7-3-1 Hongo, Bunkyo-ku, Tokyo 113-8656, Japan\\
$^4$Department of Physics, University of Massachusetts, Amherst, MA 01003, USA
}

\begin{abstract}
We study how manifestations of strong electron-phonon interaction (EPI) depend 
on the carrier concentration by solving the two-dimensional Holstein model for 
the spin-polarized fermions using an approximation free bold-line
diagrammatic Monte Carlo (BDMC) method.
We show that the strong EPI, obviously present at very small Fermion concentration, 
is masked by the Fermi blockade effects and Migdal's theorem to the extent that 
it manifests itself as moderate one at large carriers densities.
Suppression of strong EPI fingerprints is in agreement with experimental observations in doped
high temperature superconductors.
\end{abstract}

\maketitle

Discussions on the role of the EPI in the physics of cuprate compounds
with high superconducting transition temperatures (high $T_c$) have been going for decades
\cite{Anderson1997,Alexandrov1996,Anderson2007,Alexandrov2007,Gun2008,Mishchenko2009}
without resulting in a consensus opinion.
While the role of EPI in superconductivity is still under debate,
its strong manifestations were clearly observed in numerous other phenomena in high $T_c$ materials
\cite{Gun2008,Mishchenko2009,RoGu2004,tJ_Holstein_2004,RoMany2005,tJ-Tdep_2007,tJNonlocal_2007,Hub-Hol-tdep,Fausti2014,Raman}.
The apparent puzzle is that strong EPI effects seen in spectroscopic data of
undoped and weakly doped compounds become much less pronounced with hole doping
\cite{Lanzara2001,tJ_OC_2008,Carbone2008,AP2011}.
Hence, having a clear picture of how the EPI effects change with the carrier concentration is of
seminal importance for understanding the nature of unconventional superconductors where
rigorous studies are hindered by the complexity of many-body fermion problem.
Accurate results on the EPI in many-fermion systems may provide the way to reconcile
the observed fingerprints of the strong EPI in the underdoped regime with
successful descriptions of the strongly doped high $T_c$ materials by models 
based on direct electron-electron interactions alone.

More generally, it is a long standing fundamental problem to reveal how the
Migdal's theorem \cite{Migdal,Husanu2020}
emerges at the large fermion concentration and eliminates the need for vertex 
corrections even for strong EPI,
provided the Fermi-liquid state remains stable. 
The crossover between the two regimes is expected to take place at
$\omega_{\mbox{\scriptsize ph}} \sim \varepsilon_{F}$,
where $\omega_{\mbox{\scriptsize ph}}$
is the phonon frequency and $\varepsilon_{F}$ is the Fermi energy, and it can be addressed by the
approximation diagrammatic Monte Carlo methods
\cite{DMC1998,DMC2000,tJ_Holstein_2004,ManyPolaron_2014}.
To this end, we consider a spin polarized (SP) two-dimensional (2D) lattice  
system in order to avoid system instabilities that would be triggered by the 
strong EPI in continuous and spin-balanced systems, such
as structural transitions or a singlet on-site bipolaron 
formation at $\lambda \approx 0.5$ (in 2D) \cite{Macridin2004}
with the concomitant superconducting state.
An essential feature of the SP model resembling that of the $t-J$ model near half-filling
\cite{KLR1989,Dagotto1994} (which is prototypical for description of high $T_c$ superconductors)
is that in both cases one can only create one hole per site.

In this work we employ the BDMC technique developed for many-body systems 
with EPI in Ref.~\cite{ManyPolaron_2014}.
For the same system parameters the determinant Monte Carlo \cite{DM1981,DM1993} method would suffer
from a severe sign problem. The dynamical mean-filed method (DMFT) \cite{DMFT,Gunn2011},
would be inadequate because the EPI self-energy is strongly momentum dependent at 
low carrier concentration \cite{ManyPolaron_2014}, in violation of the key DMFT assumption.
The BDMC technique is based on the expansion of irreducible free-energy Feynman diagrams 
in terms of exact electron, $G$, and bare, $D^{(0)}$, phonon propagators \cite{Neglect_phonons} 
and is free from the above limitations.
In more detail, see Ref.~\cite{ManyPolaron_2014}, the electron self-energy $\Sigma^{(m)}$ 
is expanded into the series of irreducible skeleton graphs up to the largest order $m$ 
defined by the number of $D^{(0)}$ propagators, with
self-consistency implemented by a feedback loop involving the solution of the algebraic
Dyson equation,
$[G(\bf{k},\omega_\ell)]^{-1} = [G^{(0)}(\bf{k},\omega_\ell)]^{-1} - \Sigma^{(m)}(\bf{k},\omega_\ell)$,
in momentum, $\bf{k}$, and Matsubara frequency,
$\omega_{\ell} = 2 \pi T(\ell+1/2)$, representation ($\ell$ is an integer).
\begin{figure}[tbh]
\begin{center}
\includegraphics[scale=0.42]{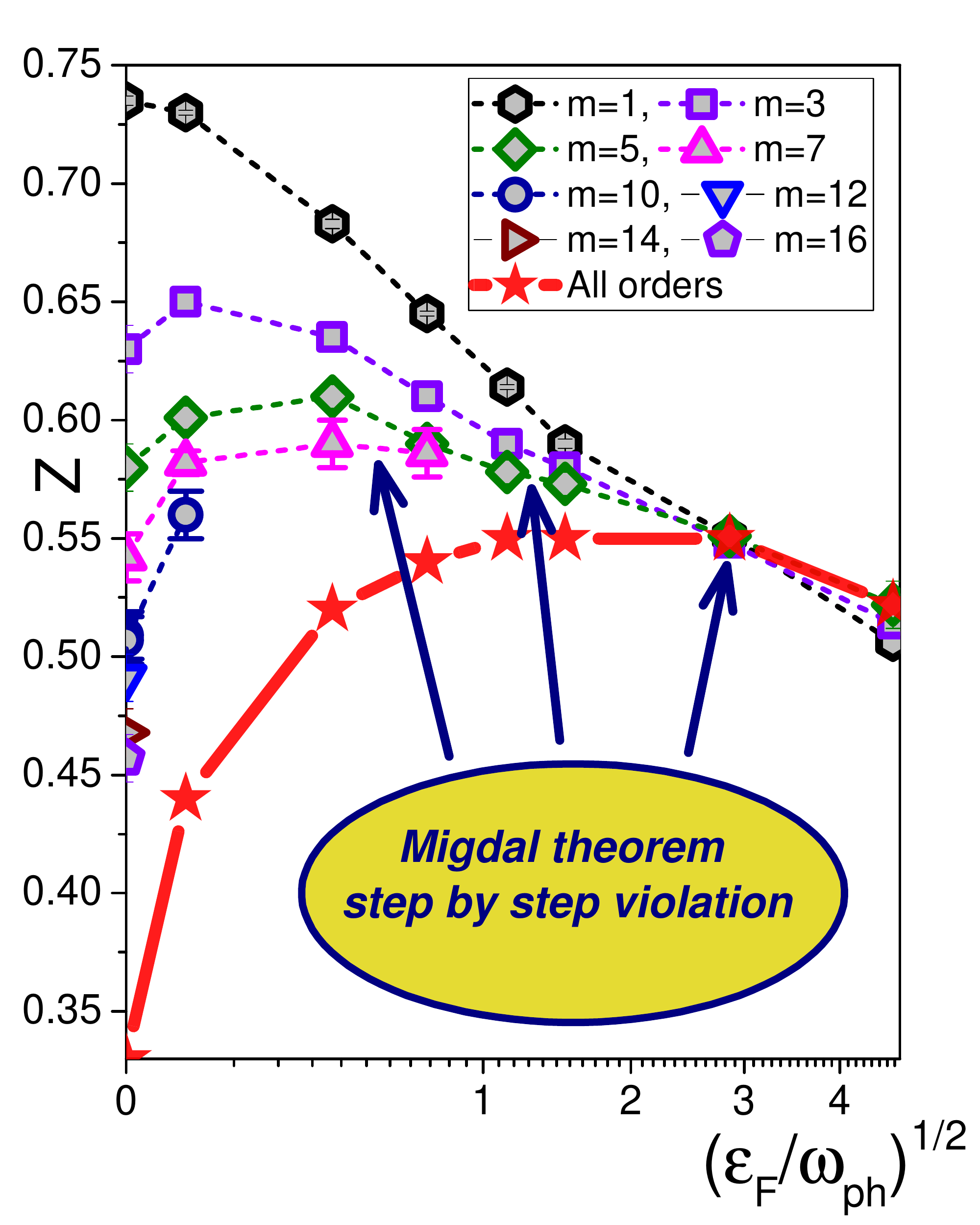}
\caption{
Quasi-particle residue at the Fermi (FS) as a function of ratio 
between the Fermi energy and phonon frequency without ($m=1$) and with vertex corrections ($m>1$).
Symbols and dashed lines represent data obtained by skeleton expansions truncated at some 
finite order $m$. 
The solid red line with stars is obtained by extrapolation to the infinite diagram-order limit $m \to \infty$.
The errorbars, if not visible, are smaller than the symbol sizes.
}
\label{fig:fig1}
\end{center}
\end{figure}

The 2D Holstein model on a square lattice reads:
\begin{equation}
H=-t \!\! \sum_{<i,j>} c^{\dagger}_{i}c_{j}^{\,}  +
\omega_{\mbox{\scriptsize ph}} \sum_{i}b_{i}^{\dagger} b_{i}^{\,} +
g \sum_{i} c_{i}^{\dagger}c_{i}^{\,} \left( b_{i}^{\dagger} +b_{i}^{\,} \right) ,
\label{h0}
\end{equation}
where $c_{i}^{\dagger}$/$b_{i}^{\dagger}$ are standard notations for electron/phonon
creation operators, $t$ is the nearest neighbor hopping amplitude,
$\omega_{\mbox{\scriptsize ph}}=0.5t$ is the energy of the local optical mode,
and $g$ is the EPI coupling.
The electron gas is spin-polarized and, hence, any site can be occupied
by no more than one electron. It is standard to characterize the strength of the EPI
by a dimensionless coupling constant $\lambda = g^2 / (4 \omega_{\mbox{\scriptsize ph}} t)$.
The lattice constant $a$, amplitude $t$, and Planck's constant $\hbar$
are used to set units of length, energy, and time, respectively.
In this study we chose $\lambda=1.07$ beyond the crossover from weak- to strong-coupling
regimes for single polarons and the threshold for the singlet bipolaron formation. 
For convenient systematic error-free handling of the data in momentum space 
we perform simulations for finite systems with $16 \times 16$ sites, large enough 
to reproduce the infinite system results with high accuracy (see Supplemental Material \cite{supplement}).
The temperature is set to $T=t/20$, which is an order of magnitude smaller than all energy 
scales of the model parameters.
In the zero-density limit an alternative exact (numerically) 
diagrammatic Monte Carlo (DMC) approach for single polarons  
\cite{DMC1998,DMC2000} provides reference values for the ground state energy, $E_1=-4.891$, 
and the quasiparticle (QP) residue, $Z_1=0.238$.

Our main results are presented in Figs. \ref{fig:fig1},  \ref{fig:fig2}, and \ref{fig:fig3}.
Figure~\ref{fig:fig1} shows the dependence of the QP residue on
the adiabaticity ratio $\gamma = \varepsilon_{F} / \omega_{\mbox{\scriptsize ph}}$.
One can see in Fig.~\ref{fig:fig1} that at large $\gamma \ge 3$
the Migdal's theorem ensures that vertex corrections are small and the lowest-order $m=1$ skeleton
diagram for self-energy (also known, depending on the context, as the non-crossing,
the self-consistent Born, and the Eliashberg approximations)
well describes the EPI renormalization even at strong coupling.
In contrast, for smaller values of  $\gamma $ one has to account for high-order vertex corrections; 
up to order $m=7$ at $\gamma=1$ and all the way to $m>20$ for $\gamma \to 0$ 
with extrapolation to the infinite diagram-order limit.
An immediate conclusion is that EPI strongly suppresses the QP residue to values smaller that 0.5
(indicative of strong coupling) only at a rather small filling factor when $ \gamma < 1$.

\begin{figure}[tbp]
\begin{center}
\includegraphics[scale=0.3]{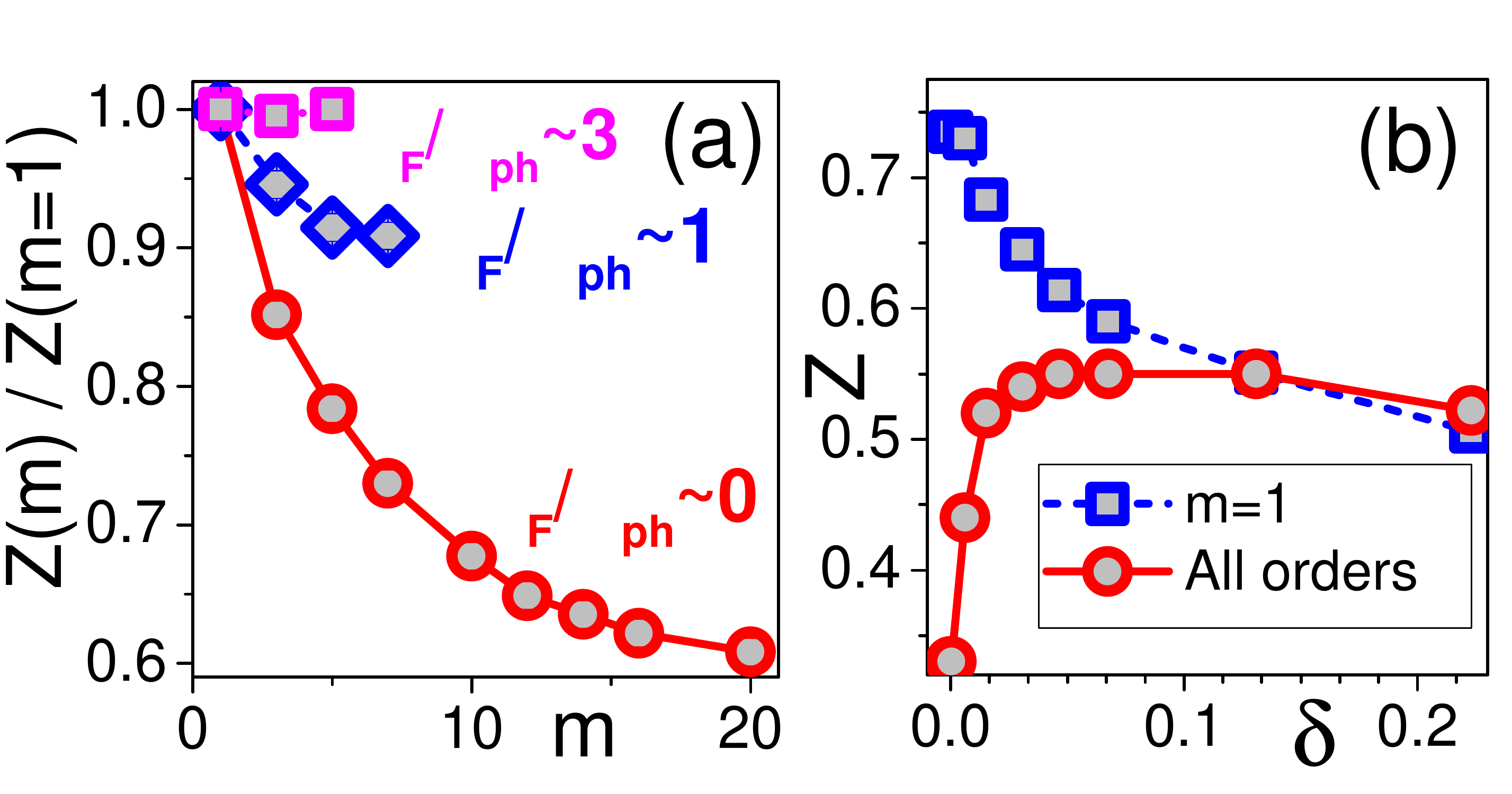}
\caption{
(a) Ratio  between the quasi-particle  residue deduced from diagrams up to order $m$
and $m=1$ (neglecting vertex corrections). Circles, diamonds, and squares stand for
$\gamma \to 0$ ($\delta=3.8 \times 10^{-4}$),
$\gamma =0.71$ ($\delta=0.0308$), and
$\gamma =2.86$ ($\delta=0.131$), respectively.
(b) Quasi-particle residue at the Fermi (FS) as a function of carrier concentration $\delta$
(circles, infinite diagram-order limit) in comparison with the $m=1$ result (squares),
see also Fig.~\ref{fig:fig1}.
}
\label{fig:fig2}
\end{center}
\end{figure}

In Fig.~\ref{fig:fig2}(a) we further quantify the role of vertex corrections at low and high carrier density (or occupation number per site), $\delta$, in both adiabatic and anti-adiabatic regimes. 
Vertex corrections become important at $\gamma <3$, and at low values of $\gamma$ and $\delta$ 
it is not sufficient to take into account just $m=2$, or even $m=3$ contributions; in this parameter 
regime the convergence is reached only for $m \ge 16$ in the skeleton expansion, see Fig.~\ref{fig:fig2}(a)).
Figure~\ref{fig:fig2}(b) is complementary to Fig.~\ref{fig:fig1} by presenting the data as a function
of the carrier concentration $\delta$ instead of $\gamma$. 
Signatures of strong EPI are observed at $\delta < 0.1$ that roughly corresponds to $\gamma \approx 1$. 
The key conclusion that clear manifestations of strong EPI are limited to small doping 
is consistent with experimental findings for high $T_c$ superconductors
 \cite{Lanzara2001,tJ_OC_2008,Carbone2008,AP2011}.

\begin{figure}[tbp]
\begin{center}
\includegraphics[scale=0.33]{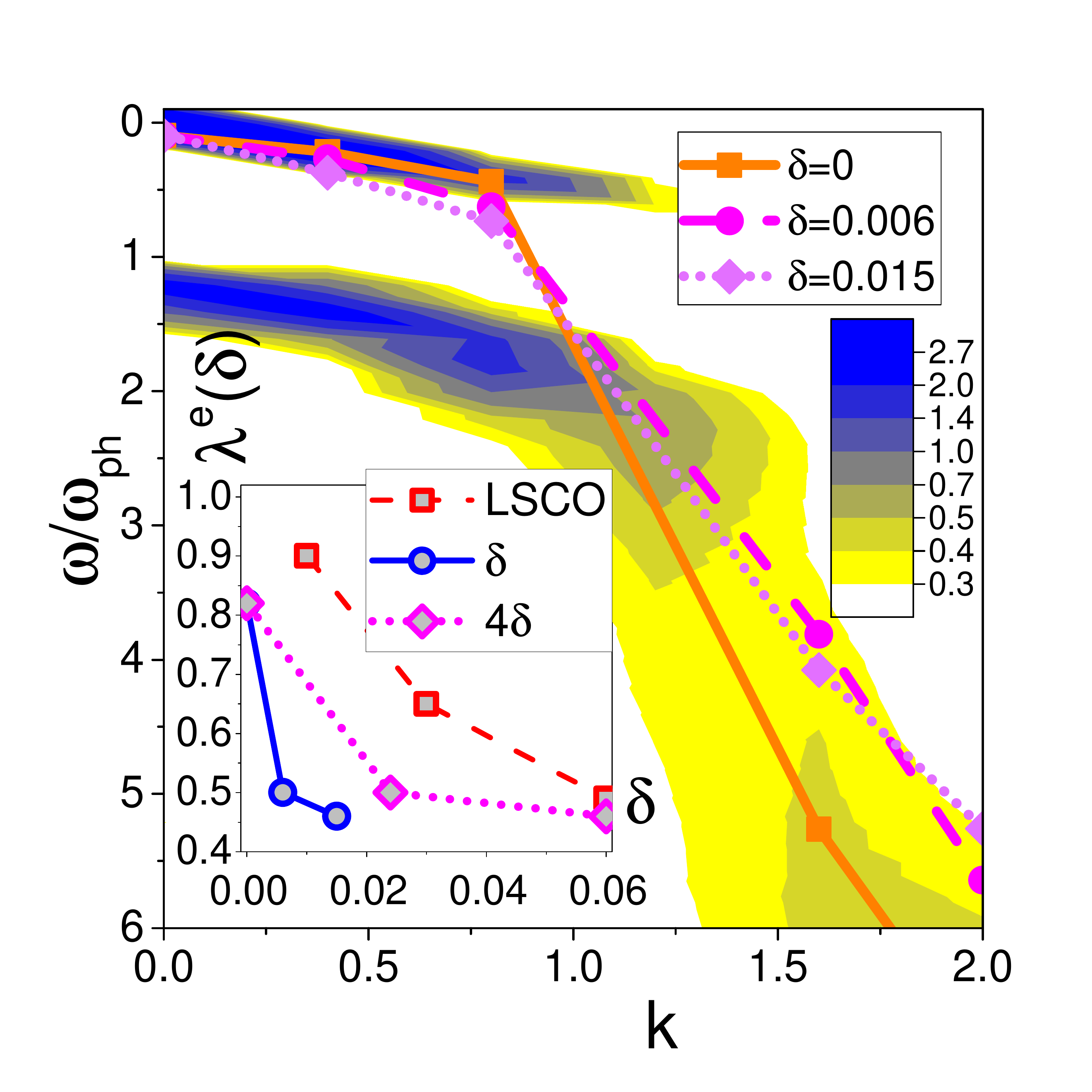}
\caption{
Contour plot of the spectral function intensity at $\delta=3.8 \times 10^{-4}$ 
with blue/yellow color used for the large/small intensity. 
Symbols connected with lines mark locations of the spectral density maxima, 
see also Fig.~\ref{fig:fig4}, for:
$\delta=3.8 \times 10^{-4}$ (squares connected by the solid line),
$\delta \approx 0.006$ (circles connected by the dashed line), and
$\delta \approx 0.015$ (diamonds connected by the dotted line).
In the inset we present the effective coupling constant $\lambda^e$ deduced from
the scaling relation (\ref{sca_lambda}) using experimental data for 
LSCO \cite{AP2011} (squares connected by a dashed line)
and locations of theoretical spectral density maxima in Fig.~\ref{fig:fig4} 
(circles connected by a solid line). We also re-plot the same theoretical data 
by using $4 \delta$ for the horizontal axis (diamonds connected by a dotted line).
Spectral densities were computed for self-energies evaluated up to 
order $m=16$ ($\delta=3.8 \times 10^{-4}$),
$m=7$ ($\delta = 0.006$), and $m=5$ ($\delta = 0.015$).
These expansion orders are enough to have converged results 
for the corresponding carrier density
(see Supplemental Material \cite{supplement}, Table I).
}
\label{fig:fig3}
\end{center}
\end{figure}

One evidence for Fermi blockade of the EPI with doping comes from
angle resolved photoemission experiments  \cite{Lanzara2001}.
It was shown that the kink angle, related to the ratio, 
$v_{\mbox{\scriptsize high}}/v_{\mbox{\scriptsize low}}$,
between the phase velocities of the dispersion relation above and below the Debye frequency,
decreases with doping. Our simulations reveal a similar trend, see Fig.~\ref{fig:fig3}.
The QP dispersion relation $\omega (\bf{ k})$ was obtained from the energy of the lowest peak
in the Lehmann spectral function, see Fig.~\ref{fig:fig4}, extracted from the 
imaginary time Matsubara Green function $G(\tau)$ by the stochastic optimization with 
consistent constraints method of analytic continuation
\cite{DMC2000,Goulko17}.

All data for the QP residues at the FS, also denoted as $Z_{FS}$, were deduced from the Fermi-liquid
relation, $Z_{FS} =[1+d]^{-1}$, with $d = \partial Re[\Sigma(k_F,\omega)]/\partial \omega \vert_{\omega=0}$.
In the low-temperature limit, the self-energy derivative at zero frequency is accurately
obtained from the ratio $-\mbox{Im}[\Sigma(k_F,\ell)]/\omega_\ell$ at the lowest Matsubara 
frequencies.
As expected, this procedure works perfectly at large carrier concentration. However,
in the zero density limit the Fermi surface shrinks to a point at zero momentum, and the entire
protocol becomes questionable. Spectral density offers an alternative way of computing the QP residue
from the integrated weight of the lowest frequency peak (we denote it as $Z_{GF}$),  see Fig.~\ref{fig:fig4}.
Somewhat surprisingly, we find that even in the zero-density limit both procedures produce consistent
results at any expansion order $m$, see inset in Fig.~\ref{fig:fig4}. At small, but finite concentration
$\delta=0.01526$ (or $\gamma =0.334$), with Fermi-momentum $k_F \approx \pi/8$ the agreement is even more precise: at order $m=5$ we find that $Z_{FS}=0.605$ and $Z_{GF}(k_F=\pi/8)=0.611$.

\begin{figure}[tbp]
\begin{center}
\includegraphics[scale=0.3]{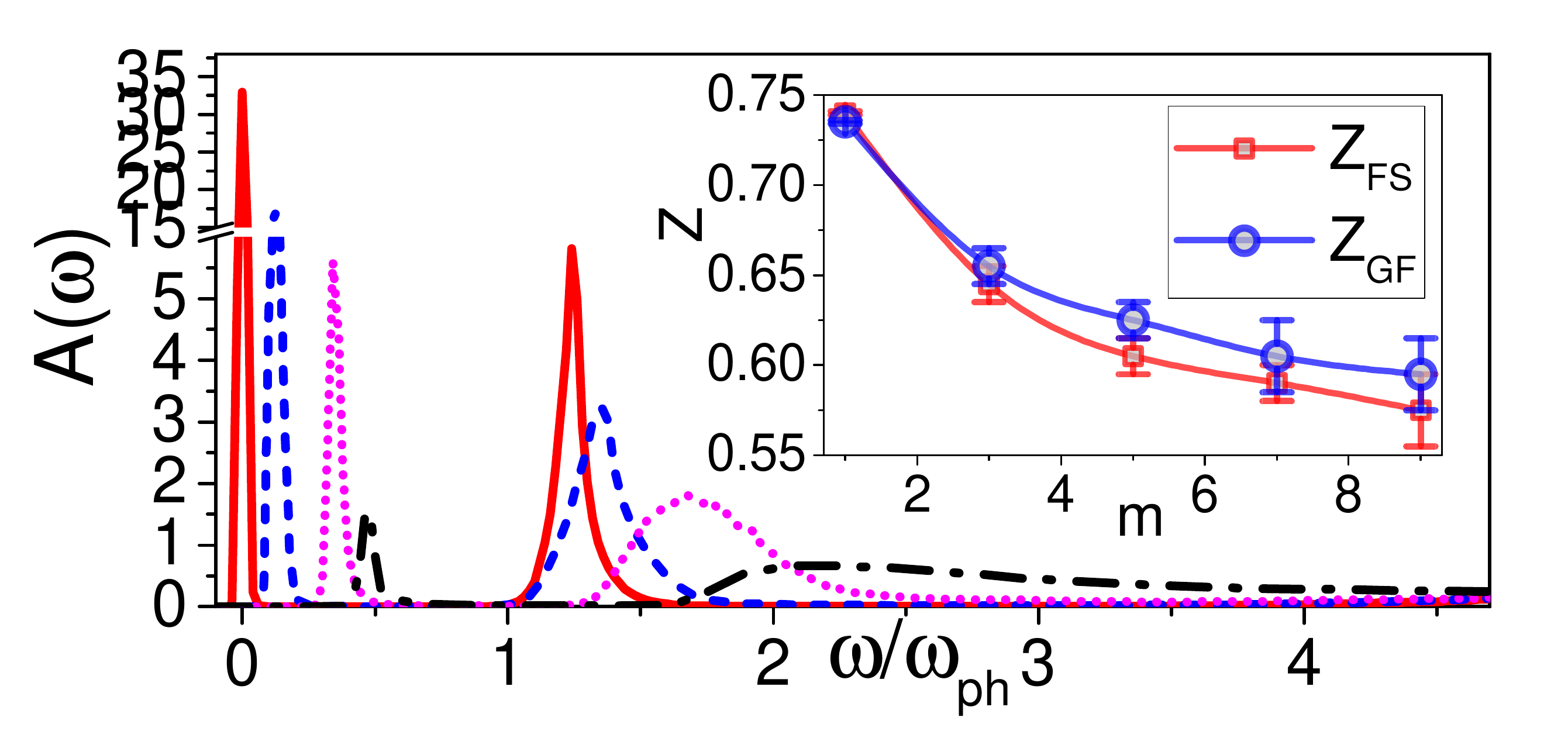}
\caption{
Spectral functions $A(\omega)$ at different momenta for $\delta =3.8 \times 10^{-4}$ from $m=16$ simulations:
$q=(0,0)$ (red solid line),  $q=(\pi/8,0)$ (blue dashed line),
$q=(2\pi/8,0)$ (magenta dotted line), and $q=(3\pi/8,0)$ (black dash-dotted line).
Energy zero was set at the value of the QP dispersion relation at $q=0$.
Inset: Order-by-order comparison between the two alternative procedures for computing
the quasi-particle residue at $q=0$:
(i) using standard Fermi liquid relations at the Fermi surface, $Z_{FS}$,
and (ii) from the lowest-frequency peak in the spectral function, $Z_{GF}$.
}
\label{fig:fig4}
\end{center}
\end{figure}

Calculations of the frequency dependent optical conductivity \cite{tJ_OC_2008} and
angle resolved photoemission spectra \cite{AP2011} in the low-concentration limit (one hole)
of the $t-J$-Holstein model revealed that the experimental dependence of both quantities on $\delta$
can be reproduced theoretically if one introduces effective EPI coupling constant
$\lambda^e(\delta)$ that decreases with doping.
It can be deduced from the photoemission spectra using scaling relation
\cite{AP2011}
\begin{equation}
\lambda^{e} = \sqrt{
\frac{v_{\mbox{\scriptsize high}}-v_{\mbox{\scriptsize low}}}{20v_{\mbox{\scriptsize low}} } \; ,
}
\label{sca_lambda}
\end{equation}
derived from nonperturbative calculations for the $t-J$-Holstein model,
where $v_{\mbox{\scriptsize low}}$ ($v_{\mbox{\scriptsize high}}$) is the velocity
above (below) the kink energy $\omega_{\mbox{\scriptsize ph}}$.
Note, the doubling of the spectral peak around the kink energy $\omega_{\mbox{\scriptsize ph}}$ is a
general feature of theoretical calculations \cite{RoGu2004,Cuk2004,Berciu2010,AP2011}.
These two peaks merge into a customary experimental picture of a single-peak kink at 
$\omega = \omega_{\mbox{\scriptsize  ph}}$ when the theoretical spectra are broadened by experimental resolution or additional damping processes \cite{Cuk2004,AP2011}. 
We compare $\lambda^e(\delta)$ deduced from experimental data of Ref.~\cite{AP2011}
with our theoretical analysis in the inset of Fig.~\ref{fig:fig3}, dashed versus solid line.
To have a meaningful quantitative comparison we also need to account for the difference
between the non-degenerate spectrum of the spin-polarized Holstein model
and fourfold degenerate ground state minimum of the experimental system. To this end we
re-plot theoretical data using $4\delta$ for the carrier concentration (dotted line).
We observe semi-quantitative agreement between the theory and experiment despite
a number of significant differences between the two cases at the microscopic level.

\begin{figure}[tbp]
\begin{center}
\includegraphics[scale=0.36]{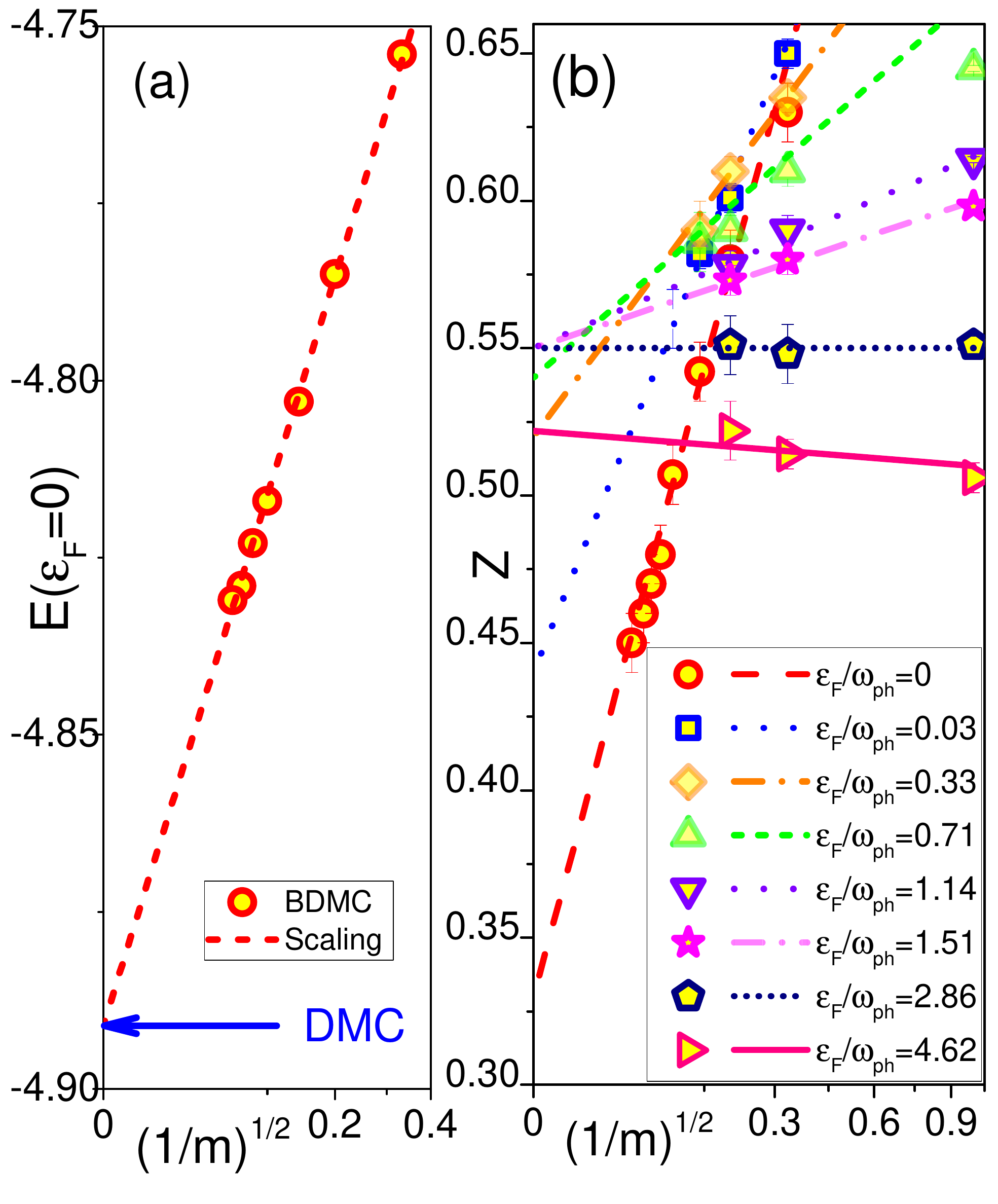}
\caption{
Finite expansion-order corrections to the polaron energy (a) 
and QP residue (b) revealing linear scaling with $m^{-1/2}$.
(a) BDMC data (circles) and the scaling law $a+b/\sqrt{m}$ (dashed line) for the
ground state energy at $\delta=3.8 \times 10^{-4}$. The DMC result at $\delta=0$ is shown by the blue arrow.
(b) BDMC data (symbols) and the scaling laws $a+b/\sqrt{m}$ (lines) for the quasi-particle residue.
}
\label{fig:fig5}
\end{center}
\end{figure}

As already mentioned in connection with Figs.~\ref{fig:fig1} and
Fig.~\ref{fig:fig2}(a), at small doping the skeleton expansion needs to go beyond $m=16$
in order to obtain correct results for the QP residue. However, both $Z$ and the polaron energy
$E$ at the FS accurately follow an empirical scaling relation, $a + b/ \sqrt{m}$,
at any carrier concentration $\delta$, see Fig.~\ref{fig:fig5}.
This allows us to perform an extrapolation to the infinite-order limit 
to eliminate the remaining systematic error as shown
in Figs.~\ref{fig:fig1}-\ref{fig:fig2}. The extrapolation procedure is validated
by an excellent agreement between the BDMC result for the ground state energy of single-polarons,
$E(m \to \infty)  =-4.89$ and the DMC benchmark $E_1=-4.891$.
The single polaron zero temperature residue $Z_1=0.238$ is renormalized  to $Z_1(\beta=20)  \approx 0.31$ 
due to finite temperature projection of the low energy self-trapping states \cite{Trugman,RashbaPekar}
(see Supplemental Material \cite{supplement}) which is also consistent with extrapolated
value $Z(m \to \infty) \approx 0.33$.

The violation of Migdal's theorem for $T=0$ is apparent in Fig.~\ref{fig:fig1} 
for all filling factors except the two largest ones.
At the lowest carrier concentrations the condition $\varepsilon_{F} \gg T$ does not 
hold any more, but this fact is barely relevant for the discussion because the theorem
is severely violated well before that, at $\varepsilon_{F} \sim \omega_{\mbox{\scriptsize ph}} \gg T$.
Thus, our finite temperature results are still valid for interpretation of the EPI suppression in
high $T_c$ materials, which is observed from low to room temperatures  \cite{Lanzara2001,tJ_OC_2008,Carbone2008,AP2011}.

\textit{Conclusions}.
We computed approximation-free results for the concentration dependence of the quasiparticle
residue $Z$ and kink angle caused by the strong electron-phonon interaction in the spin-polarized two-dimensional
Holstein model on the square lattice. We demonstrated that clear signatures of strong electron-phonon coupling
at small carrier concentration are quickly suppressed for Fermi energies exceeding the phonon frequency.
Our results provide detailed account for importance of high-order vertex corrections across the adiabatic
crossover and demonstrate that Fermi blockade of the electron-phonon interaction and irrelevance of
vertex corrections both proceed in agreement with the Migdal's theorem.
This picture explains experimental results reporting radical weakening of the electron-phonon coupling
effects in lightly doped high temperature superconductors.

{\em Acknowledgments.} N.N. and A.S.M acknowledge support by JST CREST
Grant Number JPMJCR1874, Japan, and N.P. acknowledges support of National Science Foundation
under the grant DMR-1720465 and the Simons Collaboration on the Many Electron Problem


\onecolumngrid
\appendix

\section{\large Supplementary material for ''Fermi blocade of the electron-phonon interaction:
why strong coupling effects may not be seen in optimally doped high temperature superconductors".}

Here we provide additional details on calculations performed for the 2D Holstein model
described in the main text. The lattice constant $a$, hopping amplitude $t$, and Planck's 
constant $\hbar$ are used to set units of length, energy, and time, respectively.
The phonon frequency $\omega_{\mbox{\scriptsize ph}}=0.5t$ is nearly an order of 
magnitude smaller than the particle bandwidth, and the dimensionless coupling constant
$\lambda=1.07$ corresponds to the strong coupling regime (see also below).  

\section{Size dependence}
\label{sd}

To check whether the system size $N \times N = 16 \times 16$ is sufficient to reproduce
properties of the Holstein model for single polarons when the largest finite-size effects 
are expected, we calculated various characteristics of the polaron by the 
diagrammatic Monte Carlo \cite{DMC1998,DMC2000} and compared them with known infinite lattice results.
In the simulations of finite lattice all momenta in the reciprocal space also form a
lattice  
$$k_{x,y} = (2\pi/N) j, \;\; -N/2 \le j < N/2\,.$$ 
In Fig.~\ref{fig:fig1} we show how the polaron energy, $E$, and quasiparticle
residue, $Z$, depend on the lattice size for $N = 4, 8, 16, 32, 64, 128, \infty$, and conclude
that $N=16$ results reproduce the infinite system limit with accuracy of 
three to four significant digits.

\begin{figure}[tbh]
\begin{center}
\includegraphics[scale=0.5]{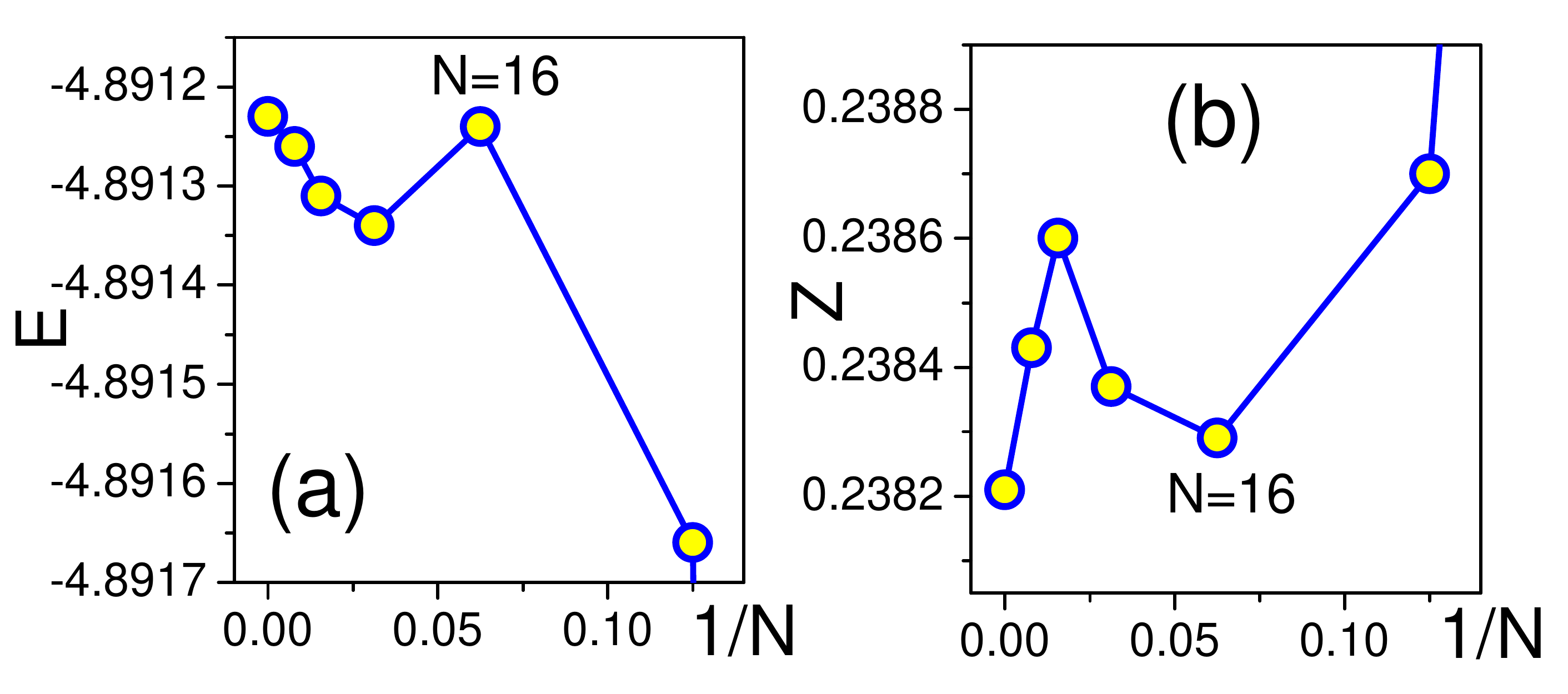}
\caption{
System size dependence of the polaron energy (a) and quasiparticle residue (b).
}
\label{fig:fig1}
\end{center}
\end{figure}

\section{Convergence of the BLDMC series as a function of carrier density}
\label{co}

\begin{figure}[tbh]
\begin{center}
\includegraphics[scale=0.5]{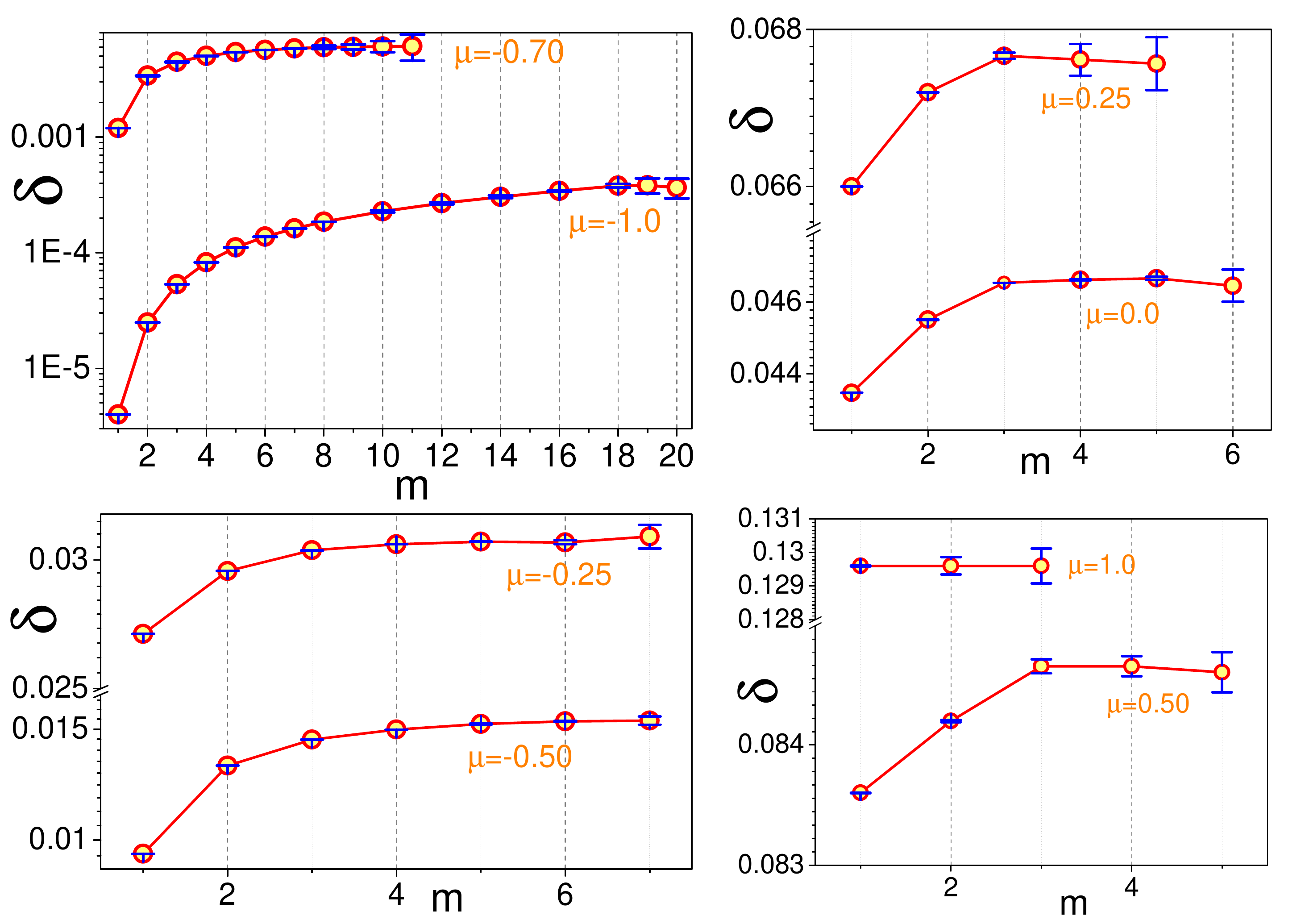}
\caption{
Fermion density dependence on the order of the self-consistent skeleton expansion
$m$ for different values of the initial chemical potential $\mu$.
}
\label{fig:fig2}
\end{center}
\end{figure}

\begin{table}[thb]
\begin{center}
\begin{tabular}{|c | c | c | c |}
\hline
$\mu$  &  $\delta$                            &     $e_F/\omega_{\mbox{\scriptsize ph}}$ &   Convergent $m$ \\
\hline
1.0     &  0.131                         &      2.86                                                      &    1 \\
0.5     & 0.085                           &     2.0                                                      &      3  \\
0.25   &  0.068                          &     1.51                                                    &       4 \\
0.0     &  0.047                           &    1.14                                                     &     4 \\
-0.25  &  0.031                        &      0.71                                                      &     4 \\
-0.5   &   0.015                        &      0.33                                                   &      5 \\
-0.7    &  0.006                          &    0.028                                                   &     7 \\
-1.0    &  $3.8 \times 10^{-4}$    &  0                                                         &      16 \\
\hline
\end{tabular}
\caption{
Relations between the chemical potential, $\mu$, fermion density per site, $\delta$, 
and ratio between the Fermi energy and phonon frequency
$e_F/\omega_{\mbox{\scriptsize ph}}$. To establish them one needs to account for 
skeleton diagrams up to order $m$.  
}
\label{tab1}
\end{center}
\end{table}

Convergence properties of the skeleton expansion strongly depend on the fermion density $\delta$ 
(or chemical potential, $\mu$, in the grand canonical ensemble). In Fig.~\ref{fig:fig2}. 
we present our BLDMC data for density dependence on the expansion order at low temperature $T = t/20$
and different values of $\mu$. At low density one needs to account for vertex corrections up to order 
$16$ to obtain converged results.
Note that the chemical potential $\mu$ is not directly related to the Fermi energy counted 
counted from the bottom of the dispersion relation which is strongly renormalized by interactions.
Table \ref{tab1} provides final relations between all quantities, including 
the required expansion order.

\section{Ground state energy, $Z$-factor, and spectral function of a single polaron}
\label{gzs}

\begin{figure}[tbh]
\begin{center}
\includegraphics[scale=0.42]{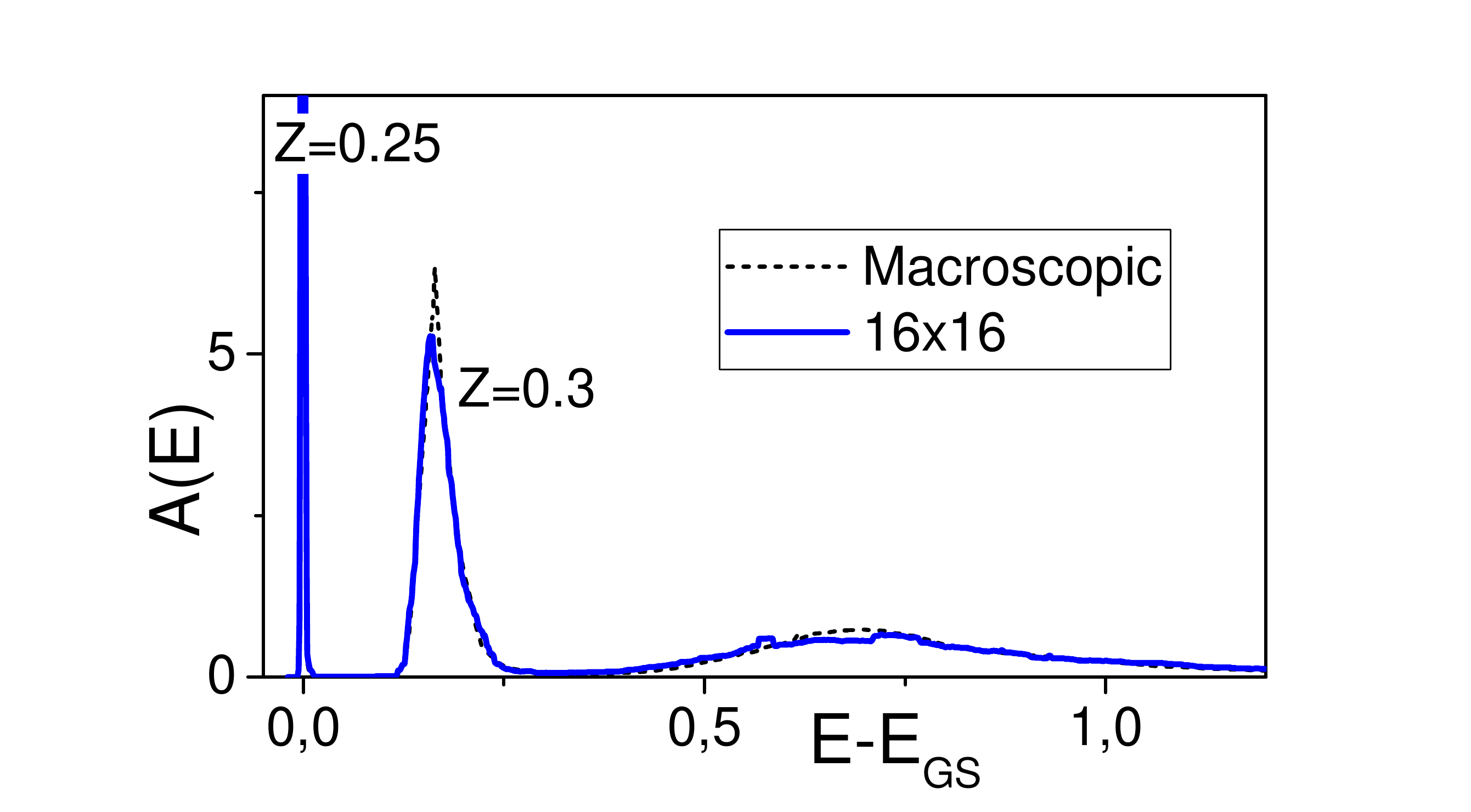}
\caption{
Spectral function of a single polaron in the infinite (black dashed line) 
and finite 16x16 (solid blue line) systems. 
}
\label{fig:fig3}
\end{center}
\end{figure}

In Fig.~\ref{fig:fig3} we present the spectral function of a single polaron in the infinite
in finite $N^2=16^2$ systems.
Nearly perfect agreement (well within the analytic continuation procedure uncertainties) 
proves that finite-size effects in this case are negligible not only for ground state
energies but also for excited states.

\begin{figure}[tbh]
\begin{center}
\includegraphics[scale=0.42]{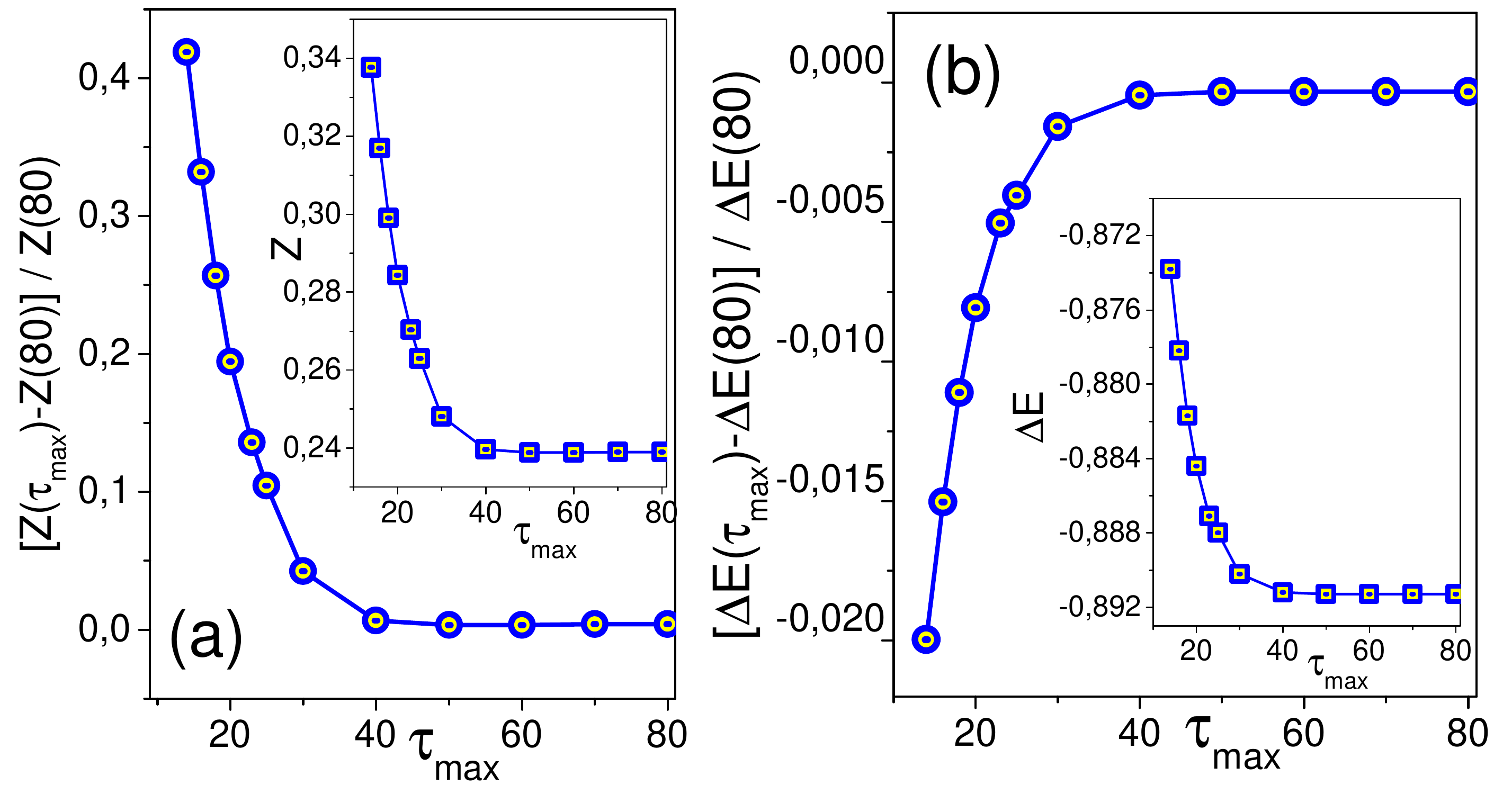}
\caption{
Saturation of estimates for the quasiparticle residue (a) and ground state energy (b)
as a function of $\tau_{max}$. For each value of $\tau_{max}$ the imaginary-time 
Green's function on the $[0.95\tau_{max},1.05\tau_{max}]$ interval was fitted by
a simple exponential function to extract $Z$ and $\Delta E$, see text.
}
\label{fig:fig4}
\end{center}
\end{figure}

To determine the quasiparticle residue and interaction induced 
energy shift, $\Delta E = E - (-4t)$, we rely on the standard reliable method:
at large imaginary time the asymptotic decay of the Green's function is given by  
$$G(\tau) \to_{\tau \to \infty} Z \exp(-\Delta E \tau) \; ,$$
see \cite{DMC1998,DMC2000}, allowing one to extract $Z$ and $\Delta E$ from a 
simple exponential fit. The leading correction decays with exponent 
controlled by the lowest excited state (the second polaron state according 
to the spectral density analysis). 
In Fig.~\ref{fig:fig4} we show how $Z$ and $E$ estimates change when we move the fitting interval $[0.95\tau_{max},1.05\tau_{max}]$
to larger values of $\tau_{max}$. It is clear from Fig.~\ref{fig:fig3} that 
the energy dependence on $\tau_{max}$ within the range $\tau_{max} \in [14,80]$ 
is very weak (about $2\%$). This is in sharp contrast, with the quasiparticle residue 
estimates: $Z$ increases by nearly 40\% when $\tau_{max}$ decreases
from $\tau_{max}=80$ to $\tau_{max}=14$.
This sensitivity explains the discrepancy between the calculations performed
at finite temperature $T=t/20$ and at $T=0$. We attribute it to the presence of
the second polaron state with comparable $Z$ factor and relatively small 
excitation energy $E_2-E_G \approx 0.17t$. 

\section{Relation of the single polaron parameters and results of extrapolation procedure for BDMC data}

 The extrapolation procedure is validated
by an excellent agreement between the BDMC result for the ground state energy of single-polarons,
$E(m \to \infty)  =-4.89$ and the DMC benchmark $E_1=-4.891$ .
In the same limit, the extrapolated result for the QP residue $Z(m \to \infty) \approx 0.33$
turns out to be larger than that for single polarons, $Z_1=0.238$.
The reason for the discrepancy is a combination of the finite temperature effect 
and self-trapping phenomenon \cite{Trugman,RashbaPekar}, manifesting itself as a second, low energy,
$E_2-E_1 \approx 0.17 <  \omega_{\mbox{\scriptsize ph}}$, excited polaron state with 
rather large spectral weight, $Z_2 \approx 0.3$,
clearly observed in the spectrum of single polarons at $T=0$, see Fig.~\ref{fig:fig3}.
Because of this soft excitation, the standard procedure of extracting $Z$  
from the large-$\tau$ asymptotic behavior of the imaginary time Green function $G(\tau)$ \cite{DMC1998,DMC2000} turns out to be sensitive to the choice of the large imaginary time 
used to fit the data (for $\tau<40$), whereas the estimate for energy remains accurate 
even for $\tau<20$, see Fig.~\ref{fig:fig4} .
Therefore, at $T=t/20$ we detect the QP weight that overestimates $Z_1$ of single polarons 
in the ground state. Semi-quantitatively, the finite temperature BDMC result can be understood
from the relation $Z_{\beta/2} = Z_1 + Z_2 \exp[-(\beta/2)(E_2-E_1)] \approx 0.31$,
which accounts for the activated second polaron state contribution at $\tau_{max}=\beta/2$.

\bibliography{letter}{}

\end{document}